\documentclass{ifacconf}

\usepackage{graphicx}      % include this line if your document 
\usepackage{natbib}        % required for bibliography
\usepackage{colortbl}
\usepackage{xcolor}
\usepackage{subcaption}
\usepackage{siunitx}
\usepackage{amsmath}
\usepackage{amsfonts}
\usepackage{amssymb}
\usepackage[inline]{enumitem}
\usepackage{mathtools}
\usepackage{algorithmicx}
\usepackage{algorithm}
\usepackage{algpseudocode}

\theoremstyle{definition}

\theoremstyle{remark}
\newtheorem{remark}{Remark}

\begin{document}
\begin{frontmatter}

\title{Deep Learning Explicit Differentiable Predictive Control Laws for Buildings} 

% TODO journal extension - JoPC
% Learning Explicit Differentiable Predictive Control of Building Thermal Dynamics
% simulations: 1 residential (NMPC), 2 office model, 3 EED dataset with learned model with disturbances as emulator

% \thanks[footnoteinfo]{ TODO .}

\author[First]{J\'an Drgo\v na} 
\author[First]{Aaron Tuor}
\author[First]{Elliott Skomski}
\author[First]{Soumya Vasisht}
\author[First]{Draguna Vrabie}

\address[First]{Pacific Northwest National Laboratory,
	Richland, Washington, USA (e-mails: 	\{jan.drgona, aaron.tuor, elliott.skomski, soumya.vasisht, draguna.vrabie\}@pnnl.gov).}

\begin{abstract}               
We present a differentiable predictive control (DPC) methodology for learning constrained control laws for unknown nonlinear systems.
DPC poses an approximate solution to multiparametric programming problems emerging from explicit nonlinear model predictive control (MPC). 
Contrary to approximate MPC, DPC does not require supervision by an expert controller. Instead, a system dynamics model is learned from the observed system's dynamics, and the neural control law is optimized offline by leveraging the differentiable closed-loop system model.
The combination of a differentiable closed-loop system and penalty methods for constraint handling of system outputs and inputs allows us to optimize the control law's parameters directly by backpropagating economic MPC loss through the learned system model.  
 The control performance of the proposed DPC method is demonstrated in simulation using learned model of multi-zone building thermal dynamics.

\end{abstract}


\end{frontmatter}
%===============================================================================

\section{Introduction}

% classical MPC for buildings
Model predictive control (MPC)~\citep{Mayne2014:MPC} is a successful control method that found its way into building control applications including thermal energy storage \citep{tang2019model}, thermal comfort \citep{park2018comprehensive} and optimization of energy management systems \citep{hil15}.  MPC with linear predictive models and constraints has been extensively studied and successfully applied over several decades for set-point tracking and economic optimization~\citep{tang2019model}. However, building models are generally nonlinear and more complex than expressed by Jacobian linearizations \citep{liu2018nmpc}. Expanding research on numerical methods and tools~\citep{Andersson2019} for optimal control problems paved the way for wider use of nonlinear MPC (NMPC) in building control applications~\citep{TACO_infrax}.
% computational issues of MPC and scalable methods
However, despite the advances in computational tools, the complexity of the online optimization is preventing low-cost deployment and maintenance of MPC in building control applications.

% explicit MPC
Explicit MPC~\citep{BemEtal:aut:02,alessio2009survey} aims to overcome the online computational drawbacks by using off-line optimization based on multiparametric programming to obtain an explicit control law, whose online evaluation is reduced to simple function evaluations~\citep{Parisio:CDC14}.
However, despite its appealing features, current multiparametric solvers~\citep{MPT3:2013} unfortunately do not scale to the complexity required by building control applications.
% approximate MPC
Addressing the scalability issues,  approximate MPC relies on learning-based methods to obtain explicit control laws by imitating MPC from observational data~\citep{DRGONA2018,GANGO2019152}.  \cite{karg2018efficient} show that deep neural networks with ReLU activations can exactly represent piecewise affine explicit MPC control laws. While \cite{hertneck2018learning} synthesize a neural network-based approximate robust MPC controller with  statistical guarantees for closed-loop stability and constraint satisfaction.
% Alternative approaches use learning methods to "warm-start" the online solvers to improve the inference speed~\citep{KLAUCO20191, chen2019large}.

% % modeling issues
% The above-mentioned methods still rely on accurate model representation of the controlled system. However, obtaining accurate building thermal dynamics models suitable for MPC typically requires expert knowledge or large datasets~\citep{AFRAM2015121}. As pointed out by~\cite{pri13} these factors limit the applicability of MPC and increase the overall cost of its implementation. 
% % RL 
% Reinforcement learning (RL)  methods have been proposed as a model-free alternative for end-to-end learning of optimal building control laws~\citep{YANG2015577}. 
% % The promise of RL methods comes from the non-reliance on expensive building modeling and direct 
% % control optimization through interactions with the environment.
% Despite its rising popularity in research, applying RL 
% in real-world applications presents several unresolved challenges~\citep{RL_challenges2019,recht2018tour}.
% For instance, RL suffers from high sample complexity, poor scalability to high dimensional problems, difficulties in off-line learning settings, and safety issues associated with inadequate constraint handling. Besides, active interaction with the environment may be infeasible when the system involves expensive hardware components. 

% differentiable MPC
A promising research direction emerged in recent years, combining constrained optimization with deep learning in so-called differentiable optimization~\citep{agrawal2019differentiating}.
% These new methods provide a framework for the systematic integration of MPC principles in learning-based controllers.
% For instance, 
\cite{diffMPC2018} learn the controller by differentiating through the KKT conditions of underlying linear MPC problem, while
\cite{east2020Infinite-Horizon} propose an infinite-horizon differentiable MPC framework based on discrete-time algebraic Riccati equation. 
\cite{chen2019gnuRL} introduced  Gnu-RL, a method that
adopts the differentiable MPC~\citep{diffMPC2018} by combining it with policy gradient algorithm for online learning.
However, differentiable MPC methods described thus far require the supervision of a pre-computed expert controller to generate the training data, similarly as in the case of approximate MPC.

In this paper, we focus on differentiable predictive control (DPC)~\citep{drgona2020differentiable}, a recently proposed learning-based method for synthesizing explicit control laws for unknown dynamical systems.
Even though similar in spirit to differentiable MPC approaches, DPC can provide scalable optimization of explicit control laws directly without the supervision of an expert controller.
Previous studies of DPC have been performed considering unknown linear~\citep{Drgona_stable}, and nonlinear~\citep{drgona2020differentiable} systems in single-input single-output (SISO) settings.
In this paper, we expand the capabilities of DPC with the following contributions:
\begin{enumerate}
    \item Use of novel system identification method for the unknown nonlinear dynamical systems, combining principles of constrained optimization with deep neural networks and state-space models.
    \item Differentiable parametrization of the closed-loop dynamics models composed of neural controllers and the identified neural state-space models.
    \item Scalable constrained deep learning of explicit control laws for multi-input multi-output nonlinear systems with economic objectives and dynamic input and output constraints enforced during learning using penalty methods.
    \item Empirical demonstration of the closed-loop control and scalability of DPC in simulations using a white-box multi-zone building model as a controlled system.
\end{enumerate}

The presented case study represents an empirical evaluation of the
DPC method for learning constrained control policies for multi-input multi-output systems.  
Dealing with the plant-model mismatch and lack of theoretical stability guarantees represent the main
limitations of the proposed study and will be addressed in future work. 

% A newer formulation called differentiable predictive control (DPC) \citep{drgona2020differentiable} mirrors MPC in its structure but drops the need for a system model to devise the control policy. The model is learned end-to-end from time-series data of the system dynamics and the policy is learned 

% NMPC allows for explicit inclusion of state and input constraints while optimizing a time-domain performance index online. Despite suffering from limited real-time applicability, recent research with enhanced models, improved stability and robustness guarantees and high performance computing capabilities have enabled

% \section{Building Modeling}

% This section introduces two types of models. A white-box building emulator used as a plant model and a physics-informed data-driven model for approximating the unknown nonlinear dynamical system used for prediction and subsequent model-based control synthesis.

\section{Differentiable Predictive Control}
\label{sec:DPC}
\begin{figure*}[htb!]
  \begin{center}
  \includegraphics[width=1.0\linewidth]{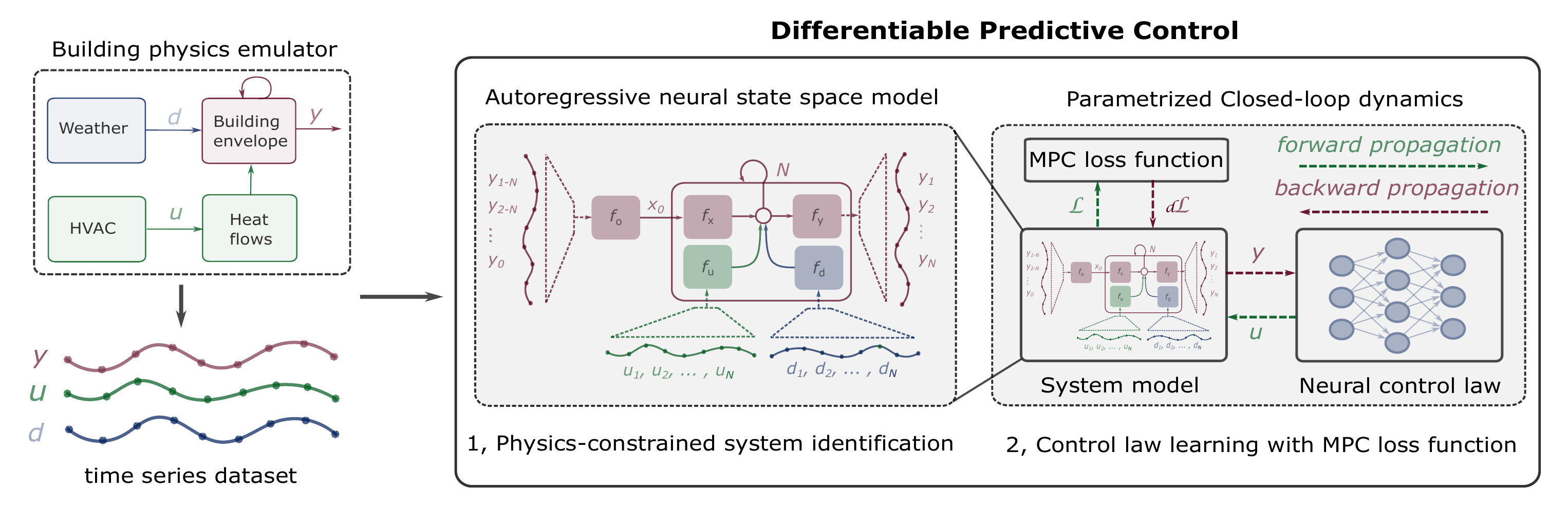}
 \end{center}
   \caption{Conceptual overview of Differentiable Predictive Control (DPC) methodology. Step 1: physics-constrained system identification.  Step 2: Learning neural control law by backpropagation of the MPC loss through the system model.}
\label{fig:DPC_method}
 \end{figure*}

In this section, we introduce Differentiable Predictive Control (DPC), a new deep learning-based constrained control method inspired by MPC.
The conceptual methodology shown in Fig.~\ref{fig:DPC_method} consists of two main steps.
Assuming time series dataset obtained from the measurements of the system dynamics, in the first step, we perform system identification using a physics-constrained  neural state-space model (SSM) introduced in section~\ref{sec:SSM}.
In the second step, we close the loop by combining the neural SSM~\eqref{eq:RNN} with neural control law obtaining a differentiable closed-loop dynamics model parametrized by neural networks.
This allows us to leverage standard tools in deep learning to optimize the neural control law parameters by backpropagating the MPC-inspired loss function through a the learned system dynamics model.

\subsection{Neural State Space Models}
\label{sec:SSM}

As a inherent part of the DPC methodology, we consider the approximation of the controlled system dynamics
by means of neural state space models (SSM)~\citep{MastiCDC2018,drgona2020physicsconstrained}.
In this paper, we consider a neural SSM architecture given as follows:
\begin{subequations}
\label{eq:RNN}
\begin{align}
\mathbf{x}_{t} &=  \mathbf{f}_o([\mathbf{y}_{t-N}; \ldots; \mathbf{y}_{t}])  \\
\mathbf{x}_{t+1} &= \mathbf{f}_{u}(\mathbf{x}_t) + \mathbf{f}_{u}(\mathbf{u}_t) + \mathbf{f}_{d}(\mathbf{d}_t) \\
\mathbf{y}_{t+1} &= \mathbf{f}_{y}(\mathbf{x}_{t+1})
\end{align}
\end{subequations}
where $\mathbf{f}_{x}$ and $\mathbf{f}_{y}$ represent the state  and output dynamics, respectively. 
 The dynamics of control actions $\mathbf{u}_t$, and disturbances $\mathbf{d}_t$ is modeled by nonlinear sub-modules $\mathbf{f}_{u}$, and $\mathbf{f}_{d}$.
 All components, $\mathbf{f}_{x}$, $\mathbf{f}_{y}$,
  $\mathbf{f}_{u}$, and $\mathbf{f}_{d}$ can be either
 parametrized by deep neural networks or linear maps. To handle partially observable systems, we use additional neural network $ \mathbf{f}_o$, representing  state observer estimating the hidden states $\mathbf{x}_t$ from the past time series of the output measurements. 
 Then combining all equations in~\eqref{eq:RNN}
gives us nonlinear autoregressive model
mapping measurements of the system outputs  over the past $N$-steps $[\mathbf{y}_{t-N}; \ldots; \mathbf{y}_{t}]$ to future time step prediction $\mathbf{y}_{t+1}$.

\subsection{Neural Parametrization of the Closed-Loop Dynamics}
Constructing a differentiable model of the  closed-loop dynamical system  represents a core conceptual idea behind the DPC methodology.
In this paper, we use the neural state space model~\eqref{eq:RNN}  to represent the open-loop system dynamics and a fully connected neural network control law $\boldsymbol{\pi}_{\theta}(\boldsymbol {\xi}): \mathbb{R}^m \rightarrow \mathbb{R}^n$ 
given as:
\begin{subequations}
    \label{eq:dnn}
    \begin{align}
   \boldsymbol{\pi}_{\theta}(\boldsymbol {\xi}) & =  \mathbf{W}_{L}  \mathbf{h}_L + \mathbf{b}_{L} \\
    \mathbf{h}_{l} &= \boldsymbol\sigma(\mathbf{W}_{l-1} \mathbf{h}_{l-1} + \mathbf{b}_{l-1})  \label{eq:dnn:layer}\\
    \mathbf{h}_0 &= \boldsymbol {\xi}
 \end{align}
\end{subequations}
where $\boldsymbol{\pi}_{\theta}(\boldsymbol {\xi})$  parametrized by
$\boldsymbol \theta = \{\mathbf{W}_1, \ldots \mathbf{W}_{L}, 
\mathbf{b}_1, \ldots \mathbf{b}_{L}\}$
with $\mathbf{W}_{l}$ and  $\mathbf{b}_{l}$  
representing weights and biases
for hidden layers $l \in \mathbb{N}_1^L$, respectively. 
Then  $\boldsymbol\sigma: \mathbb{R}^{n_h} \rightarrow \mathbb{R}^{n_h}$ represents the elementwise application of a univariate activation function $\sigma: \mathbb{R} \rightarrow \mathbb{R}$ to linearly transformed hidden layer states $ \mathbf{h}_l$ as given in~\eqref{eq:dnn:layer}.

In this paper, the control law  $\mathbf{U}_f  = \boldsymbol{\pi}_{\theta}(\boldsymbol {\xi})$ is mapping the
features $\boldsymbol{\xi}$ to future control actions trajectories $\mathbf{U}_f = [\mathbf{u}_{t}; \ldots; \mathbf{u}_{t+N}]$ where $N$ defines the length of the prediction horizon.
The computed control trajectories $\mathbf{U}_f$ are used to rollout the system dynamics model~\eqref{eq:RNN} over $N$-steps ahead into the future.
 Then we close the loop by
 using past output trajectories of the system dynamics  $\mathbf{Y}_p = [\mathbf{y}_{t-N}; \ldots; \mathbf{y}_{t}]$ as features $\boldsymbol \xi = \mathbf{Y}_p$ for the the neural control law  $\boldsymbol{\pi}_{\theta}(\mathbf{\xi})$.
 For more expressive control laws, the 
features $\boldsymbol{\xi}$ can contain additional vectors such as disturbance forecasts $\mathbf{D}_f$, or previews of dynamic references and constraints imposed on states, inputs or outputs.  
For the sake of brevity, we compactly represent the closed-loop dynamics  as follows:
\begin{subequations}
    \label{eq:closedloop}
    \begin{align}
  \mathbf{U}_f & =  \boldsymbol{\pi}_{\theta}(\boldsymbol \xi)  \\
    \mathbf{Y}_f & =  \mathbf{f}_{\text{SSM}}([\mathbf{Y}_p;  \mathbf{U}_f; \mathbf{D}_f]), 
 \end{align}
\end{subequations}
where $\boldsymbol{\pi}_{\theta}$ gives the control law \eqref{eq:dnn}
and $\mathbf{f}_{\text{SSM}}$ represents $N$-step ahead rollout of the system dynamics model~\eqref{eq:RNN}.
The parametrization of the closed-loop model~\eqref{eq:closedloop} by deep neural networks (DNN) is motivated by the expressivity and scalability of DNNs.
% , as well as by practical aspects of scalable and user-friendly software packages. 
% In principle, deep neural networks could be replaced by any differentiable multivariate functions. 
% Hence, the DPC methodology  is not tight to a particular parametrization of the closed-loop dynamics model.

\begin{remark}
% The presented parametrization of the closed-loop dynamics model~\eqref{eq:closedloop} could be modified in various ways.
To decrease the dimensionality of the feature space, we could leverage the latent space $\mathbf{x}_{t}$ of the prediction model $\mathbf{f}_{\text{SSM}}$ as policy features.
Similarly, we could apply the principle of receding horizon control (RHC) to the DPC policy $\boldsymbol{\pi}_{\theta}$, computing only the first time step
of the control action $\mathbf{u}_{t}$ instead of a  control trajectory $\mathbf{U}_f$.
\end{remark}

\subsection{Constraints Satisfaction via Penalty Methods}

%  data-driven model-free control methods became popular in recent years.
% Despite their rising popularity, the constraints satisfaction represents a significant challenge
%  in most of the model-free methods present in the literature.

Constraints satisfaction is a premier feature of MPC facilitated by system dynamics equations. 
We argue that incorporating  principles of MPC 
in the parametrized closed-loop model
represents a systematic way of handling constraints in learning-based control synthesis.
In particular, having explicit parametrizations of the system dynamics~\eqref{eq:RNN} and control law~\eqref{eq:dnn}
allow us to simulate the influence of the computed control actions on future state trajectories. 
% This parametrization allows us to leverage a wide variety of constrained optimization methods.
In this work, we enforce the time-varying constraints 
on the control actions and system outputs 
by using penalty functions:
\begin{subequations}
\label{eq:penalty}
\begin{align}
& p(\mathbf{x}_t, \mathbf{\overline{x}}_t) = 
\text{max}(0,\:\mathbf{x}_t - \mathbf{\overline{x}}_t) 
\:\:\: &\cong \:\: \mathbf{x}_t - \mathbf{s}^{\overline{x}}_t \leq \mathbf{\overline{x}}_t
 \label{eq:penalty:ub} \\
& p(\mathbf{x}_t, \mathbf{\underline{x}}_t) = \text{max}(0,\:-\mathbf{x}_t + \mathbf{\underline{x}}_t)
\:\:\: &\cong \:\:
\mathbf{\underline{x}}_t \leq \mathbf{x}_t + \mathbf{s}^{\underline{x}}_t   \label{eq:penalty:lb} 
\end{align}
\end{subequations}
The penalties quantifying the constraint violations are incorporated in to the loss function of the learning problem.
In the context of deep learning, they can be straightforwardly implemented using standard \texttt{ReLU} functions.
% The utility of the penalty methods in optimal control has been validated for decades~\citep{SNYMAN199247}. 
In this paper, we leverage more recent use of penalties to impose constraints in the context of deep learning~\citep{MarquezNeilaSF17}.
This allows us  to enforce the inequality constraints
 during optimization of the closed-loop dynamics model~\eqref{eq:closedloop} parametrized by neural networks. 
 
 \begin{remark}
 Even though in this paper, the authors use \texttt{ReLU}  functions to model the penalties~\eqref{eq:penalty}, other commonly used continuously differentiable functions such as \texttt{GELU}, \texttt{ELU}, or \texttt{Softplus} could be used as well.
 \end{remark}

\subsection{Economic MPC Loss Function}

The problems in the classical process control applications are typically formulated as reference tracking subject to constraints. In building control applications, instead of tracking setpoint values, we are typically concerned with economic objectives such as energy use or cost minimization while maintaining the desired thermal comfort level prescribed by temperature constraints. 
In classical control literature, we refer to this method as economic MPC~\citep{ELLIS20141156,Rawlings6425822}.
To achieve the desired performance of the closed-loop dynamics~\eqref{eq:closedloop}, we formulate the following loss functions inspired by economic MPC:
\begin{equation}
\label{eq:loss}
\begin{split}
\mathcal{L} = \frac{1}{nN} \sum_{i=1}^{n}  \sum_{k=1}^{N} \Big(
 Q_{\text{umin}}||{\bf u}^i_t||^2_2 +
 Q_{\text{du}}||{\bf u}^i_t - {\bf u}^i_{t-1}||^2_2 +
\\ Q_{\text{y}}||p({{\bf y}}^i_t, {\underline{{\bf y}}}^i_t)||^2_2 +
Q_{\text{y}}||p({{\bf y}}^i_t, {\overline{{\bf y}}}^i_t)||^2_2 + \\ Q_{\text{u}}||p({{\bf u}}^i_t, {\underline{{\bf u}}}^i_t)||^2_2 +
Q_{\text{u}}||p({{\bf u}}^i_t, {\overline{{\bf u}}}^i_t)||^2_2 \Big)
    \end{split}
\end{equation}
Here $t$ is a time index of $N$-step ahead prediction horizon, and $i$ is an index of a total number of $n$ batches of the training time-series data. 
The first term of the loss function weighted with $ Q_{\text{umin}}$ represents the energy minimization term by pushing the control actions towards zero values. The second term weighted with $ Q_{\text{du}}$ penalizes the control rate of change
and represents a simple strategy for smoothening the control trajectories. Third and fourth terms weighted with $Q_{\text{y}}$ are constraints penalties on controlled outputs, while the last two terms weighted with $Q_{\text{u}}$ are penalties on control action constraints. 
Using economic MPC objective with soft constraints penalties offers a principled way of optimizing the parametrized closed-loop dynamics~\eqref{eq:closedloop} with complex performance criteria.

\subsection{Optimization of Differentiable Predictive Control Laws}

In this section, we present DPC as a two-step algorithm (Algorithm~\ref{algo:DPC})
for optimizing explicit control laws using parametrized differentiable closed-loop dynamics model~\eqref{eq:closedloop}.
\begin{algorithm}[ht!]
  \caption{Differentiable Predictive Control (DPC)}\label{algo:DPC}
  \begin{algorithmic}[1] 
   \State  \textbf{Inputs:} datasets for modeling, $\boldsymbol D_{\text{sysID}}$, and control $\boldsymbol D_{\text{ctrl}}$.
   \State   \textbf{Step 1:} System identification of the model $\mathbf{f}_{\text{SSM}}$~\eqref{eq:RNN} using the dataset $\boldsymbol D_{\text{sysID}}$ of observed system dynamics.
    \State   \textbf{Step 2:}  Learning explicit control law $\boldsymbol{\pi}_{\theta}$~\eqref{eq:dnn} by optimizing economic MPC loss function~\eqref{eq:loss} subject to closed-loop dynamics model~\eqref{eq:closedloop} 
    using synthetic dataset $\boldsymbol D_{\text{ctrl}}$.
    \State  \textbf{Outputs:}  Learned system dynamics  model $\mathbf{f}_{\text{SSM}}$~\eqref{eq:RNN} and explicit control law $\boldsymbol{\pi}_{\theta}$~\eqref{eq:dnn}.
  \end{algorithmic}
\end{algorithm}

The first step of the DPC algorithm~\eqref{algo:DPC}
performs system identification of the neural state-space model~\eqref{eq:RNN}. 
In the second step of Algorithm~\ref{algo:DPC}, we optimize the parameters ${\theta}$ of  neural control law $\boldsymbol{\pi}_{\theta}$~\eqref{eq:dnn} by
minimizing the loss function $\mathcal{L}$~\eqref{eq:loss} subject to 
 closed-loop dynamics model~\eqref{eq:closedloop} and penalty constraints on states and outputs~\eqref{eq:penalty}, respectively.
As part of the closed-loop model~\eqref{eq:closedloop}, we use the system dynamics model $\boldsymbol M$~\eqref{eq:RNN} with parameters obtained from the system identification in step one. 

For decoupling the system identification from control optimization, the system model parameters in~\eqref{eq:RNN}  remain fixed during the optimization of the control law parameters~\eqref{eq:dnn} using the backpropagation algorithm. In the forward pass, we
sample the features $\boldsymbol {\xi}$ of the control law~\eqref{eq:dnn} to generate candidate control trajectories $\mathbf{U}_f$ and rollout the closed-loop dynamics $N$-steps ahead into the future to obtain the output response $\mathbf{Y}_f$ of the system dynamics model~\eqref{eq:RNN}.  
Then we compute the derivatives of the loss function~\eqref{eq:loss} and backpropagate them through the closed-loop dynamics model~\eqref{eq:closedloop} to optimize the parameters of the neural control law~\eqref{eq:dnn} using  stochastic gradient descent updates.

From a control-theoretic perspective, the structure of the unrolled closed-loop system dynamics model~\eqref{eq:closedloop} in DPC resembles the structure of the constraints in the dense MPC formulation, also called the single shooting approach. 
Then the optimization of the DPC control policy 
via automatic differentiation 
of differentiable closed-loop system model~\eqref{eq:closedloop} can be interpreted as 
an offline model-based iterative learning approach.
As a result, DPC is capable of synthesizing highly complex constrained control laws
by optimizing the simulated closed-loop behavior without solving supervisory MPC as in the case of imitation learning~\citep{hertneck2018learning,Zhang2019SafeAN}.

\section{Simulation Case Studies}

\subsection{Emulator of Building Physics}
\label{sec:emulator}

As a ground truth plant model, we use a white-box building thermal model developed using Modelica's \texttt{IDEAS} library \citep{IDEAS} and envelope linearization method~\citep{picard2015linearization}.
The building of interest represents a six-zone residential house located in Belgium. The heating ventilation and air conditioning  (HVAC)
system consists of a central gas-boiler and a single radiator per zone of the building. 
From the control perspective, the system has six outputs representing zone operative temperatures, 
 seven control actions corresponding to one supply temperature of the central heating unit, and six mass flow rates, one per zone.  The building envelope is modeled by $286$ unmeasured states, and the dynamics are additively affected by $41$
 environmental disturbances such as solar irradiation and ambient temperature. For system identification and control, we assume to have the forecast of only the ambient temperature.
This model was previously used in the MPC context in~\cite{DRGONA2018}. We refer to this reference  for further technical details on the emulator's building physics.

\subsection{Datasets and Optimization}

For the system identification step of DPC algorithm~\ref{algo:DPC}, we use the building physics emulator described in section~\ref{sec:emulator} to generate the time series dataset $\boldsymbol{D}_{\text{sysID}}$ sampled with $T_s = 15$ min rate,
in the form of tuples of control inputs $\mathbf{u}$,   disturbance signals $\mathbf{d}$, and system outputs $\mathbf{y}$ given as:
\begin{equation}
    \label{eq:dataset}
    \begin{split}
   \boldsymbol D_{\text{sysID}} = \{(\mathbf{u}^{(i)}_t, \mathbf{d}^{(i)}_t, \mathbf{y}^{(i)}_t), (\mathbf{u}^{(i)}_{t+T_s}, \mathbf{d}^{(i)}_{t+T_s}, \mathbf{y}^{(i)}_{t+T_s}),  \ldots, \\
    (\mathbf{u}^{(i)}_{t+NT_s}, \mathbf{d}^{(i)}_{t+NT_s}, \mathbf{y}^{(i)}_{t+NT_s})\},
    \end{split}
\end{equation}
with $n$ batches indexed by $i = \mathbb{N}_1^n$  of time series trajectories, each $N$-steps long. 
We use this dataset that carries the building operation data to train the parameters of physics-constrained autoregressive state-space model~\eqref{eq:RNN}. 
 
In the second step of DPC algorithm~\ref{algo:DPC},
we train the control law $\boldsymbol{\pi}_{\theta}(\boldsymbol {\xi})$ using following synthetically generated training dataset:
\begin{equation}
    \label{eq:dataset_ctrl}
    \begin{split}
   \boldsymbol D_{\text{ctrl}} = \{(\tilde{\mathbf{y}}^{(i)}_{t-NT_s},
   \underline{{\bf y}}_t^{(i)}, \overline{{\bf y}}_t^{(i)},
     \underline{{\bf u}}_t^{(i)}, \overline{{\bf u}}_t^{(i)}, \mathbf{d}^{(i)}_t
   ),  \ldots, \\
    (\tilde{\mathbf{y}}_{t}^{(i)},
   \underline{{\bf y}}_{t+NT_s}^{(i)}, \overline{{\bf y}}_{t+NT_s}^{(i)},
     \underline{{\bf u}}_{t+NT_s}^{(i)}, \overline{{\bf u}}_{t+NT_s}^{(i)}, \mathbf{d}^{(i)}_{t+NT_s}
   )\},
    \end{split}
\end{equation}
where $\tilde{\mathbf{y}}$ are sampled past output trajectories, which represent perturbation of the
initial conditions for the closed-loop system dynamics~\eqref{eq:closedloop}.
Similarly we sample $\underline{{\bf y}}$, $\overline{{\bf y}}$, $ \underline{{\bf u}}$, and $\overline{{\bf u}}$ representing  time-varying lower and upper bounds on controlled outputs and control actions, respectively.
For measured disturbances $\mathbf{d}$ we assume to have access to their $N$-step ahead forecast.
We select a subset of trajectories from dataset $\boldsymbol{D}_{\text{ctrl}}$  to act as features $\boldsymbol {\xi}$ for the neural control law~\eqref{eq:dnn}
with their compact notation
$\tilde{\mathbf{Y}}_p = [\tilde{\mathbf{y}}^{(i)}_{t-NT_s}; \ldots; \tilde{\mathbf{y}}^{(i)}_{t}]$, 
$\underline{\mathbf{Y}}_f = [\underline{{\bf y}}^{(i)}_t; \ldots; \underline{{\bf y}}^{(i)}_{t+NT_s}]$,
${\mathbf{D}}_f = [ \mathbf{d}^{(i)}_t; \ldots; \mathbf{d}^{(i)}_{t+NT_s}]$.
Both datasets $\boldsymbol{D}_{\text{sysID}}$ and $\boldsymbol{D}_{\text{ctrl}}$ contain $8640$ time samples corresponding to  $90$ days of the building operation data. We split each to training, development, and test sets of equal length with $2858$ time samples.
 
 We implement the DPC algorithm~\ref{algo:DPC} using Pytorch library~\citep{paszke2019pytorch}.
 In the first step,
we train the system model~\eqref{eq:RNN} with $25266$ parameters.
Subsequently, we construct the closed-loop dynamics model~\eqref{eq:closedloop} and train the neural control law~\eqref{eq:dnn} over the prediction horizon of $N=32$ steps by optimizing the loss function~\eqref{eq:loss}
using the Adam optimizer \citep{kingma2014adam}, with $1000$ gradient descent updates, randomly initialized weights, and learning rate of $0.001$,  
  The resulting control law  $\boldsymbol{\pi}_{\theta}(\boldsymbol {\xi}): \mathbb{R}^{416} \rightarrow \mathbb{R}^{224}$ 
     with $4$-layers with $100$ hidden units and GELU activation functions
  maps $416$ features $\boldsymbol {\xi} =[\tilde{\mathbf{Y}}_p; \underline{\mathbf{Y}}_f; \mathbf{D}_f]$ to $224$ control action points $\mathbf{U}_f$ what corresponds to $188624$ parameters in total.
 Optimizing this control law using DPC algorithm~\ref{algo:DPC} and the above-specified setup took under $100$ seconds on a desktop machine 
 with  $64$-bit $2.60$ GHz Intel(R) i7-8850H CPU and $16$ GB RAM.

\subsection{Physics-constrained System Identification}

To improve the accuracy and generalization of the    proposed neural state space model~\eqref{eq:RNN} we constrain the model based on known building physics. 
The white-box building models typically use thermal resistance-capacitance (RC) networks and convective heat flow equations as summarized in~\cite{drgona2020physicsconstrained}.
These type of building models can be accurately approximated by 
Hammerstein architecture, with state  $\mathbf{f}_{x}(\mathbf{x}_{t}) = \mathbf{A}\mathbf{x}_{t}$, and  output $\mathbf{f}_{y}(\mathbf{x}_{t}) = \mathbf{C}\mathbf{x}_{t}$ dynamics represented by linear maps.
Then the system dynamics matrix $\mathbf{A}$ approximates the RC network describing the thermal dynamics of the building envelope structure, $\mathbf{f}_{u}$
is the nonlinear HVAC dynamics approximating the convective heat flow equation, and  $\mathbf{f}_{d}$ is the nonlinear effect generated by the ambient conditions.
The system outputs 
 $\mathbf{y}_{t}$ represent zone operative temperatures, 
the hidden states $\mathbf{x}_{t}$ represent lumped temperatures of the building envelope, control actions $\mathbf{u}_{t}$ denote the supply temperature and mass flows for each zone, while disturbance signal $\mathbf{d}_{t}$
represents the ambient temperature.
This interpretation allows us to enforce physically realistic bounds on the dynamic behavior of individual components. 
For instance, we know that
building envelopes are dissipative systems. Thus we impose constraints on the eigenvalues of $\mathbf{A}$ to learn a strictly stable system~\citep{tuor2020constrained}. 
Additionally, we can use the penalty method to constrain the input dynamics influence to remain within a realistic range.

\subsection{Closed-Loop Control}

\begin{figure*}[htb!]
  \begin{center}
  \includegraphics[width=0.9\linewidth]{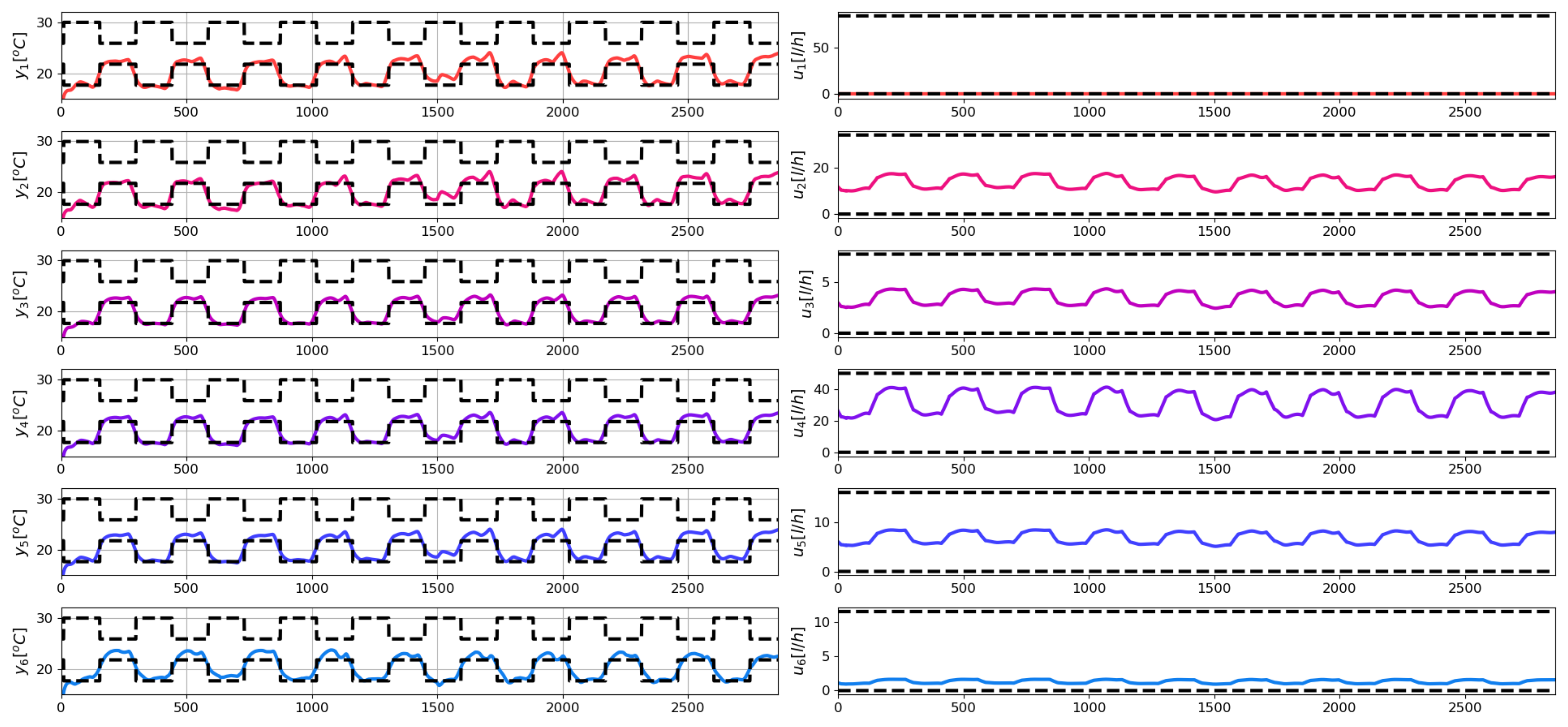}
 \end{center}
   \caption{Closed-loop control performance of explicit DPC control law evaluated on learned nominal model. Left column represents controlled zone temperatures. Right column shows manipulated mass flows.}
\label{fig:DPC_nominal}
 \end{figure*}

We demonstrate the control performance 
of the explicit control laws obtained by means of DPC algorithm~\ref{algo:DPC}.
Fig.~\ref{fig:DPC_nominal} plots the closed-loop control trajectories evaluated on the learned state-space model~\eqref{eq:RNN}, hence considering no plant-model  mismatch. The left column represents controlled outputs (zone temperatures) that need to stay within time-varying bounds.
The right column shows constrained control actions (mass flows) for each output computed by explicit DPC control laws.
These trajectories demonstrate the capability to synthesize control laws with near-optimal performance in terms of economic objectives and constraints satisfaction using DPC algorithm~\ref{algo:DPC}. 
However, when deployed to control the emulator, the 
 DPC policy learned on the nominal model was not able to compensate for the plant model mismatch. 
Thus these initial results indicate that
due to close-to-optimal performance on the nominal model (Fig.~\ref{fig:DPC_nominal}), the field performance of the DPC boils down to a problem of accurately identifying the system dynamics with coupling.

\section{Limitations and Future Work}

The lack of theoretical stability guarantees and the ability to deal with the plant model are the main limitations of the DPC methodology presented in this study. 
%  Without the use of the  PI  feedback on top of the nominal DPC policy, the resulting control trajectories would violate the imposed output constraints.  
In future work, we plan to expand the DPC methodology for the systematic handling of plant-model mismatch.  
To introduce real-time feedback capabilities into DPC, we aim to incorporate the principles of robust and off-set free MPC, as well as adaptive control updates.
The authors are also working on theoretical closed-loop stability guarantees for DPC.
Our future empirical work will include a systematic comparison with classical MPC methods on experiments including a wide range of nonlinear systems.

\section{Conclusions}

This paper applied a recently proposed differentiable predictive control (DPC) methodology to a multi-zone building control problem. 
We have expanded the prior capabilities of DPC in learning complex explicit control laws for unknown nonlinear dynamical systems. In particular, we have shown that DPC can systematically handle economic objectives subject to dynamic constraints imposed on nonlinear system dynamics with multiple inputs and outputs. 
The DPC closed-loop control capabilities are empirically demonstrated in a simulation case study using a multi-zone building thermal dynamics model.
We have shown that it is possible to learn explicit control laws for constrained nonlinear optimal control problems with a large number of states and long prediction horizons.
By doing so, we tackle the complexity limitations of classical explicit MPC based on multiparametric programming.
Simultaneously, DPC overcomes the main limitation of imitation learning-based approaches such as approximate MPC
by alleviating the need for supervision from the model predictive controller (MPC).
Based on the presented features, we believe that the proposed DPC methodology has long-term potential, not only for research but also for practical applications, such as building control that requires fast development and low-cost deployment and maintenance on hardware with limited computational resources.
However, several practical challenges, such as dealing with the plant model mismatch, need to be addressed in future work.

\begin{ack}

This work was funded by the Physics Informed Machine Learning (PIML) investment at the Pacific Northwest National Laboratory (PNNL). 

\end{ack}

\bibliography{main}             % bib file to produce the 

% \appendix
% \section{Appendix A}    % Each appendix must have a short title.

\end{document}